\documentclass[aps,prb,twocolumn,superscriptaddress,nobibnotes]{revtex4-1}

\usepackage{rotating}
\usepackage{amssymb,amsmath}
\usepackage{color}
\usepackage{braket}

\newcommand{\srsl}{Sr$_2$IrO$_4$}
\newcommand{\srbl}{Sr$_3$Ir$_2$O$_7$}

\begin{document}

\title{Evidence of quantum dimer excitations in Sr$_3$Ir$_2$O$_7$}

\author{M. Moretti Sala}
\thanks{Both authors contributed equally.}
\affiliation{European Synchrotron Radiation Facility, BP 220, F-38043 Grenoble
Cedex, France}

\author{V. Schnells}
\thanks{Both authors contributed equally.}
\affiliation{Institute for Theoretical Physics and Astrophysics, University of
W\"{u}rzburg, Am Hubland D-97074 W\"{u}rzburg, Germany}

\author{S. Boseggia}
\affiliation{London Centre for Nanotechnology and Department of Physics and
Astronomy, University College London, London WC1E 6BT, UK}
\affiliation{Diamond Light Source Ltd, Diamond House, Harwell Science and
Innovation Campus, Didcot, Oxfordshire OX11 0DE, UK}

\author{L. Simonelli}
\affiliation{European Synchrotron Radiation Facility, BP 220, F-38043 Grenoble
Cedex, France}
\affiliation{CELLS-ALBA Synchrotron Radiation Facility, Carretera BP 1413, km
3.3 08290 Cerdanyola del Valles, Barcelona, Spain}

\author{A. Al-Zein}
\affiliation{European Synchrotron Radiation Facility, BP 220, F-38043 Grenoble
Cedex, France}

\author{J. G. Vale}
\affiliation{London Centre for Nanotechnology and Department of Physics and
Astronomy, University College London, London WC1E 6BT, UK}

\author{L. Paolasini}
\affiliation{European Synchrotron Radiation Facility, BP 220, F-38043 Grenoble
Cedex, France}

\author{E. C. Hunter}
\affiliation{Centre for Science at Extreme Conditions, Peter Guthrie Tait Road,
King's Buildings, Edinburgh. EH9 3FD. United Kingdom}

\author{R. S. Perry}
\affiliation{London Centre for Nanotechnology and Department of Physics and
Astronomy, University College London, London WC1E 6BT, UK}

\author{D. Prabhakaran}
\affiliation{Clarendon Laboratory, Department of Physics, University of Oxford,
Parks Road, Oxford, OX1 3PU, United Kingdom}

\author{A. T. Boothroyd}
\affiliation{Clarendon Laboratory, Department of Physics, University of Oxford,
Parks Road, Oxford, OX1 3PU, United Kingdom}

\author{M. Krisch}
\affiliation{European Synchrotron Radiation Facility, BP 220, F-38043 Grenoble
Cedex, France}

\author{G. Monaco}
\affiliation{European Synchrotron Radiation Facility, BP 220, F-38043 Grenoble
Cedex, France}
\affiliation{Dipartimento di Fisica, Universit\`a di Trento, via Sommarive 14,
38123 Povo (TN), Italy}

\author{H. M. R\o nnow}
\affiliation{Laboratory for Quantum Magnetism, \'Ecole Polytechnique
F\'ed\'erale de Lausanne (EPFL), CH-1015 Lausanne, Switzerland}
\affiliation{Neutron Science Laboratory, Institute for Solid State Physics,
University of Tokyo, Kashiwa, Chiba 277-8581, Japan}

\author{D. F. McMorrow}
\affiliation{London Centre for Nanotechnology and Department of Physics and
Astronomy, University College London, London WC1E 6BT, UK}

\author{F. Mila}
\affiliation{Institute of Theoretical Physics, \'Ecole Polytechnique
F\'ed\'erale de Lausanne (EPFL), CH-1015 Lausanne, Switzerland}

\date{\today}

\begin{abstract}

The magnetic excitation spectrum in the bilayer iridate \srbl\ has been investigated using high-resolution 
resonant inelastic x-ray scattering (RIXS) performed at the iridium L$_3$ edge and theoretical techniques. 
A study of the systematic dependence of the RIXS spectrum on the orientation of the wavevector transfer, $\mathbf{Q}$, with respect to the iridium-oxide bilayer has revealed that the magnon dispersion is comprised 
of two branches well separated in energy and gapped across the entire Brillouin zone.  
Our results contrast with those of an earlier study which reported the existence of a single dominant branch.
While these earlier results were interpreted as two overlapping modes within a spin-wave model of weakly coupled iridium-oxide planes, our results are more reminiscent of those expected for a system of weakly coupled dimers.
In this latter approach the lower and higher energy modes find a natural explanation as those corresponding to transverse and longitudinal fluctuations, respectively. We have therefore developed a bond-operator theory 
which describes the magnetic dispersion in \srbl\ in terms of quantum dimer excitations. In our model dimerisation is produced by the leading Heisenberg exchange, $J_c$, which couples iridium ions in adjacent planes of the bilayer. The Hamiltonian also includes in plane exchange, $J$, as well as further neighbour couplings and relevant anisotropies. The bond-operator theory provides an excellent account of the dispersion of both modes, while the measured $\mathbf{Q}$ dependence of the RIXS intensities is in reasonable qualitative accord with the spin-spin correlation function calculated from the theory. We discuss our results in the context of the quantum criticality of bilayer dimer systems in the presence of anisotropic interactions derived from strong spin-orbit coupling.

\end{abstract}

\maketitle

\section{Introduction}

The physics of iridium-based $5d$ transition-metal oxides has triggered
considerable interest, as it represents the opportunity to explore the
consequences of electron correlations in the strong spin-orbit coupling limit.
Various novel electronic states, topological and otherwise, have been predicted
for the iridates\cite{Pesin2010,Witczak-Krempa2014}. The most studied example to
date is the relativistic Mott-like insulating state observed in \srsl\ which, in
the absence of spin-orbit coupling, would be expected to be a metal. The
insulating state has been argued to result from the combined action of strong
crystal-field and spin-orbit coupling, leading to band narrowing and a
$j_{\mathrm{eff}}=1/2$ ground state, on which electron correlations then act to
produce a charge energy gap $\Delta E$~\cite{Kim2008}.

\begin{figure}[!ht]
	\centering
		\includegraphics[width=\columnwidth]{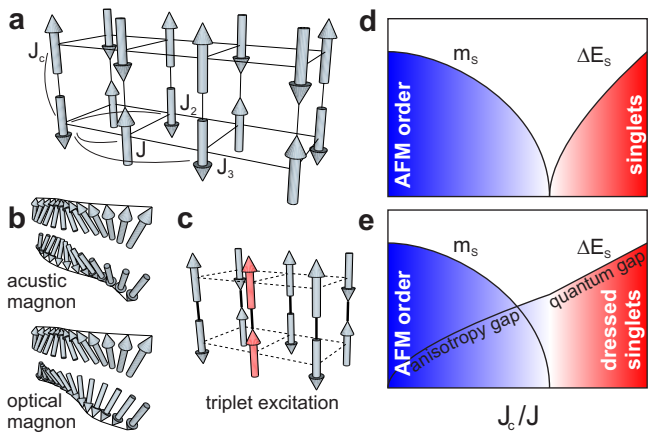}
\caption{(a)-(c) Magnetic structure and excitations in \srbl. (a) intra and
inter-layer couplings (see Eq. 1); (b) conventional acoustic and optical
spin-wave modes previously used to interpret RIXS data from \srbl\
\cite{Kim2012}; (c) dimer triplet excitation. (d)-(e) Schematic evolution of 
a bilayer system as a function of inter- to intra-layer coupling. (d) Quantum
phase transition of the SU(2) symmetric model; (e) Effect of anisotropy leading to a cross over between antiferromagnetic order and quantum dimer
regime with a gap $\Delta E_S$ that never closes for finite coupling ratios.}\label{fig1}
\end{figure}

More generally, studies of the Ruddlesden-Popper series,
Sr$_{n+1}$Ir$_{n}$O$_{3n+1}$, are proving to be particularly fruitful, as
members of this series display a striking evolution in their electronic and
magnetic properties as the number of IrO$_{2}$-layers is increased. This can be
illustrated by considering the end members of this series. Single layer \srsl
($n=1$) is an insulator ($\Delta E\approx 615$~meV~\cite{Nichols2014})
with a canted, basal-plane antiferromagnetic (AF) ground state~\cite{Kim2009,BoseggiaBP}, displaying a magnon dispersion~\cite{Kim2012c} consistent with an effective Hamiltonian dominated by Heisenberg interactions, in agreement with theory~\cite{Ge2011}. SrIrO$_{3}$ ($n=\infty$) is a strongly correlated metal with ferromagnetic correlations~\cite{Zhao2008}. Bilayer \srbl\ ($n=2$) represents the interesting case of being a marginal insulator ($\Delta E\approx 130$~meV~\cite{Okada2013}), with a $c$-axis antiferromagnetic ground
state (Fig.\ \ref{fig1}(a))~\cite{Boseggia2012,Kim2012b}, and a magnetic
spectrum dominated by a single excitation with an unusually large magnon energy
gap $\Delta E_S \sim$ 85~meV~\cite{Kim2012}.

This excitation has been interpreted as two overlapping bilayer spin-wave
modes (Fig.\ \ref{fig1}(b)) in the presence of enhanced interlayer
pseudo-dipolar coupling~\cite{Kim2012}. However, in bilayer systems, the
presence of a single dominant magnon branch is typical of weakly coupled dimers,
in which case anisotropy generically gives rise to two gapped modes close to
each other with significantly different intensities. Therefore, we reexamine the
nature of the low-energy dynamics in \srbl\ by performing high-resolution resonant inelastic X-ray scattering (RIXS) experiments which exploit a different experimental geometry compared to Ref.~\onlinecite{Kim2012}. Our experiment establishes a fundamentally different picture of the magnon dispersion in \srbl, with the observation of two distinct gapped modes. The dispersion and intensity of these modes are hard to reconcile with a spin-wave description, but they can be well accounted for by a bond-operator mean-field description that captures the quantum dimer nature of the excitations (Fig.\ \ref{fig1}(c)).

\section{Experimental}

RIXS is a photon in--photon out technique for the investigation of the electronic structure of materials by probing excitations of various nature\cite{Ament2011}. The scattering process can be described as in the following: a monochromatic photon is resonantly absorbed by the system, promoting an electron from a core level into the valence band. This state, usually referred to as the intermediate state of the RIXS process, is highly unstable and therefore short-lived. In RIXS one monitors the recombination of the core-hole to a final state of lower energy by a radiative transition. Analysis of the energy and momentum of the emitted photon allows one to characterise the final state of the RIXS process; this can be either the ground state itself, as in elastic scattering, in which case the emitted photon energy coincides with the inident one, or an excited state. In the latter case, the energy of the excited state is determined by the difference between the incident and emitted photon energy. Beside energy, momentum transfer is also used to label excitations. This is particularly informative when studying dispersive excitations, like magnetic one.

RIXS experiments were performed on the ID20 beamline of the European Synchrotron
Radiation Facility (ESRF), Grenoble, with an overall energy resolution of
25~meV. This is achieved by monochromatising the incident photons with a Si(844) back-scattering channel-cut and using a Rowland spectrometer equipped with Si(844) spherical (R=2 m) diced crystal analysers~\cite{Moretti2013}. The scattering plane and incident photon polarisation were both horizontal in the laboratory frame, i.e. $\pi$ incident polarization was used. The \srbl\ single crystal was grown by flux method of Ref.~\onlinecite{Boseggia2012}. The sample was cooled to a temperature of 15~K in a closed flow He-cryostat equipped with Be-windows. RIXS spectra were recorded with the incident photon energy fixed at 11.217 keV, approximately $3$~eV below the main absorption line as it is known that the intensity of magnetic excitations is maximized at this energy~\cite{Liu2012,Moretti2014c,Moretti2014b}. This shift provides a rough estimate of the cubic crystal-field splitting in Ir 5$d$ states consistent with previous results~\cite{Moretti2014}.

\begin{figure}
  \centering
  \includegraphics[width=.85\columnwidth]{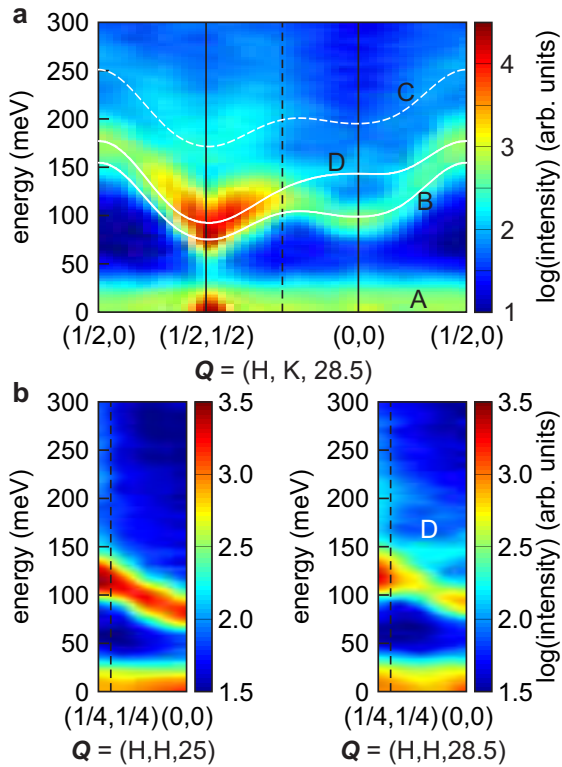}
  \caption{In-plane wavevector dependence of the RIXS response of
\srbl\ in the 0-300 meV energy range. (a) High-symmetry directions,
$\mathbf{Q}=(H,K,28.5)$ (r.l.u.). Letters A-D label modes discussed in the
text. White lines represent our model. (b) Close-up on limited region of the
Brillouin zone, $\mathbf{Q}=(H,H,25)$ (left) and $\mathbf{Q}=(H,H,28.5)$
(right).}\label{fig2}
\end{figure}

In Fig.\ \ref{fig2}(a), we present the in-plane momentum dependence of the RIXS
response from \srbl\ along high-symmetry directions of the Brillouin zone for
$L=28.5$. An elastic line (possibly containing contributions from phonons) and a
magnetic excitation dominate the spectra in the 0$-$300 meV energy range.
Following the convention in Ref. \onlinecite{Kim2012}, we label these features A
and B, respectively. The latter has a sizeable dispersion of $85\pm 5$~meV, and
a gap of comparable magnitude. A weaker feature, C, follows the dispersion of
feature B at higher energies. These observations are in good accord with
previous RIXS measurements~\cite{Kim2012}. However, closer inspection of the
dispersion along the symmetry line from (0,0) to (1/4,1/4), reveals the presence
of an additional, previously unreported feature, labelled D. This is most
prominent around (0,0) where it is clearly separated from B and C. Away from
(0,0), it merges almost into feature B and contributes to its lineshape and
spectral weight. The intensity of feature D was found to be strongly dependent
on the out-of-plane component of $\mathbf{Q}$: it completely vanishes when
changing $L$ from 28.5 to 25, as shown in the two panels of Fig.\
\ref{fig2}(b).

\begin{figure}
	\centering
		\includegraphics[width=.85\columnwidth]{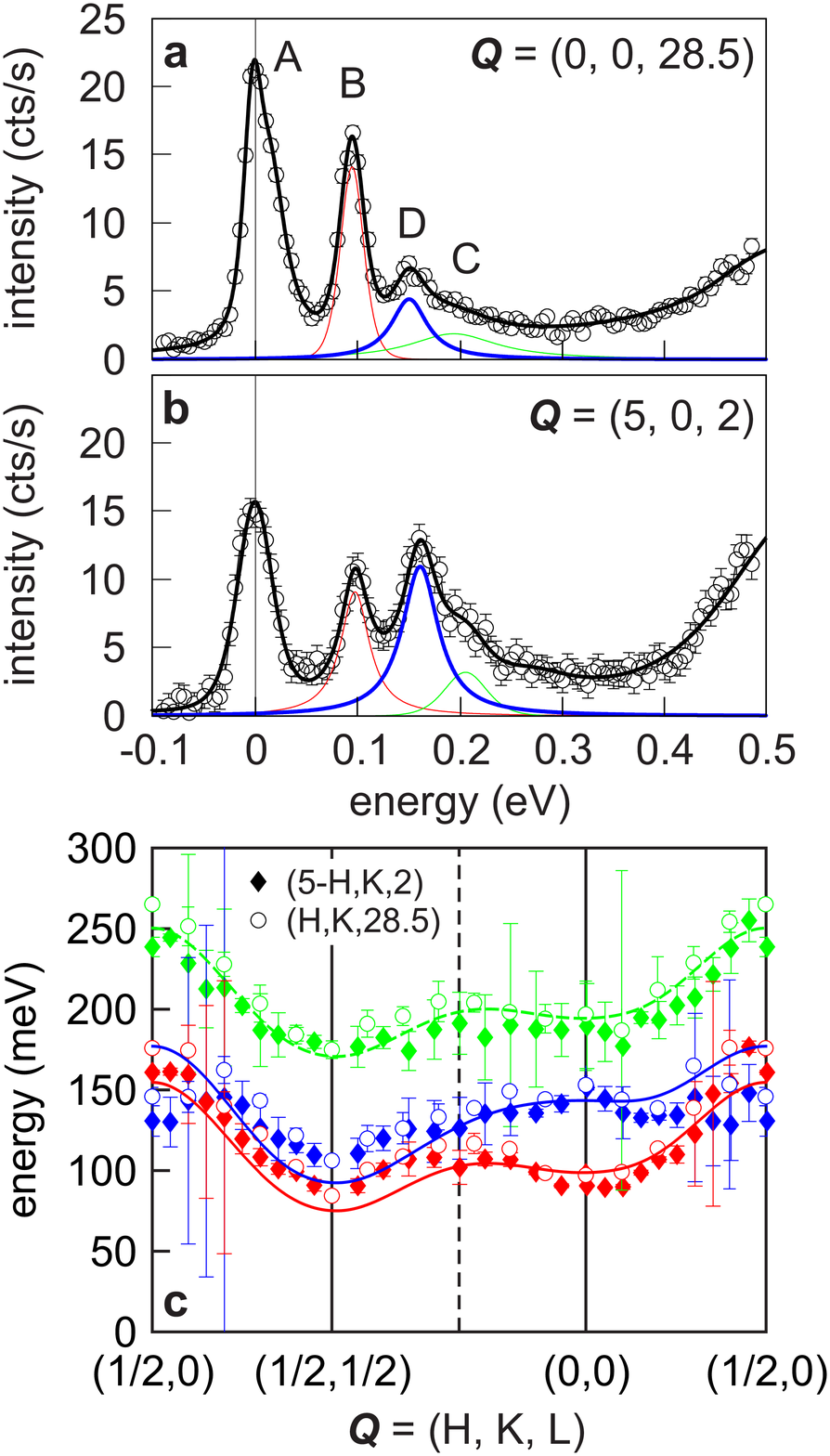}
\caption{(a)-(b) RIXS spectra from \srbl\ for (a) $\mathbf{Q}=(0, 0, 28.5)$
and (b) $\mathbf{Q}=(5, 0, 2)$. (c) Energies of features B (red), C (green) and
D (blue) as a function of in-plane momentum transfer. Experimental data: open
circles, (H, K, 28.5), filled diamonds, (5-H,K,2). Theory: continuous lines
represent the transverse (red) and longitudinal (blue) modes, respectively. As
in Fig.\ \ref{fig2}(a), the dashed line is a guide to the eye obtained by
shifting the theoretical curve for feature B by approximately 85
meV.}\label{fig3}
\end{figure}

\begin{figure}
  \centering
  \includegraphics[width=.85\columnwidth]{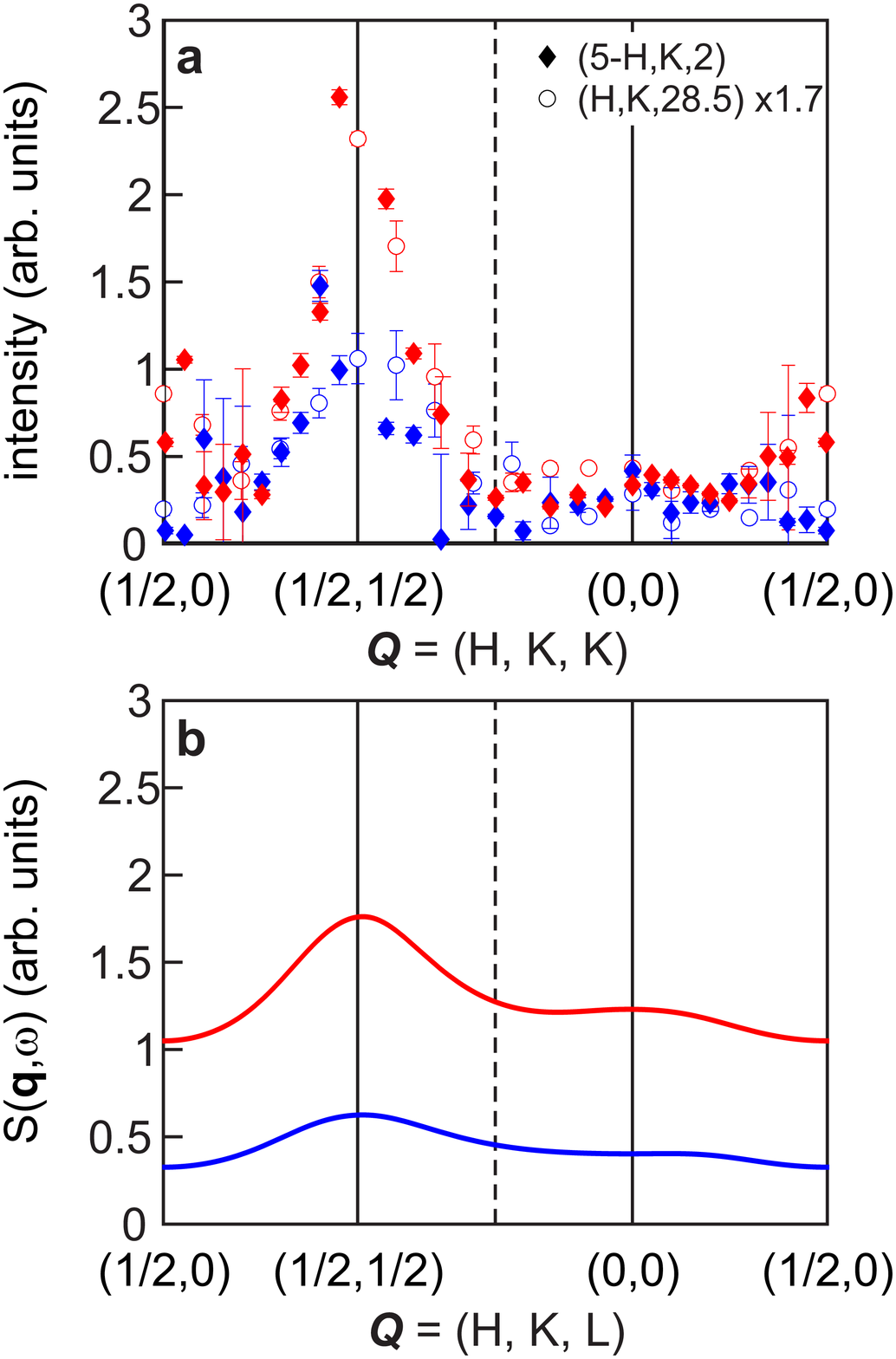}
  \caption{Integrated intensity of features B and D as a function of
in-plane momentum transfer. (a) Experimental data for features B (red) and D
(blue): open circles, (H, K, 28.5), filled diamonds, (5-H,K,2). (b) Theory:
transverse (red) and longitudinal (blue) responses.}
  \label{fig4}
\end{figure}

The spectrum corresponding to $\mathbf{Q}=(0,0,28.5)$ is displayed in
Fig.\ \ref{fig3}(a). Features B, C and D are fitted to three Pearson VII
functions~\cite{Wang2005}. Feature D is clearly visible, although its integrated
intensity is only a fraction of that of B and comparable to that of C. The
extracted dispersions of features B, C and D as a function of the in-plane
momentum transfer for $L=28.5$ are plotted in Fig.\ \ref{fig3}(c). The
corresponding wavevector dependences of the integrated intensities are shown in
Fig.\ \ref{fig4}(a).

The results discussed so far were obtained in a geometry with the wavevector
transfer $\mathbf{Q}$ predominantly perpendicular to the IrO$_2$ planes. To
explore how the RIXS spectrum depends on the orientation of $\mathbf{Q}$,
experiments were also performed with $\mathbf{Q}$ predominatly oriented
in the IrO$_2$ planes. An example of data taken in this geometry is shown in
Fig.\ \ref{fig3}(b). It is immediately clear that the relative peak intensities
of the features have a strong dependence on the orientation of $\mathbf{Q}$,
with feature D acquiring spectral weight at the expense of feature B as
$\mathbf{Q}$ is rotated towards the planes. The energies and intensities
extracted from data taken with $\mathbf{Q}$ in-plane are also plotted in Fig.\
\ref{fig3}(c) and Fig.\ \ref{fig4}(a), respectively, where they are seen to be
in good agreement with data acquired using the initial geometry.

Features B and C have already been identified and discussed in
Ref.~\onlinecite{Kim2012}. Feature B was interpreted as the superposition of almost
degenerate acoustic and optical magnons (Fig.\ \ref{fig1}(b)), and their
dispersion was modelled on the basis of a Hamiltonian which includes intra- and
interlayer couplings, as well as pseudo-dipolar and Dzyaloshinsky-Moriya
interaction terms. Feature C was assigned to the onset of a multi-magnon
continuum. Feature D was not reported in the previous experiment, most likely
because of the nontrivial dependence of its intensity on momentum transfer both
as a function of $L$ (Fig.\ \ref{fig2}(b)) and the orientation of $\mathbf{Q}$
(Fig.\ \ref{fig3}(a) and (b)) revealed here.

\section{Theory}

Let us now turn to a theoretical discussion of the magnetic excitations of \srbl.
In a SU(2) symmetric (Heisenberg) bilayer system, there is a quantum phase
transition as a function of the ratio of interlayer to intra-layer
coupling~\cite{Hida1992,Monien1993,Monien1994,Sandvik1994, Yoshioka1996,
Sommer2001,Hamer2012} between a gapless, magnetically ordered phase, and
a gapped phase (see Fig.\ref{fig1}(d)). In the limits of very weak or very strong inter-layer coupling,
linear spin-wave theory and perturbation theory starting from isolated dimers
provide very accurate descriptions respectively. However, these approaches fail in the
intermediate regime, and the only simple approach that provides a qualitatively
correct description throughout is the bond-operator mean-field
theory~\cite{Sachdev1990,Gopalan1994,Sushkov1998,Sommer2001,Bouillot2011}.

In the presence of strong anisotropy, as is the case in \srbl,
the excitation spectrum is always gapped (Fig.\ \ref{fig1}(e)), and it is impossible
to know just from the excitation spectrum in which regime the system lies. However, on
general grounds the
excitation spectrum of \srbl\ revealed by RIXS is incompatible with linear spin-wave
theory. Indeed linear spin-wave theory predicts two modes, which have
dispersions related to each other by
$\omega_{\mathrm{acustic}}(\mathbf{q})=\omega_{\mathrm{optical}}((1/2,
1/2)-\mathbf { q } )$ ($\mathbf{q}$ is the in-plane momentum transfer), implying
that the spectrum should be symmetric around $(1/4,1/4)$ in Fig.\ \ref{fig3}(c),
which is clearly not the case.

We therefore developed a description of \srbl\ in terms of
coupled dimers. In this approach, the parameter that controls the center of the
main band is the interlayer coupling $J_c$, which must then be of the order of
100 meV. The fact that the dispersion is approximately degenerate at (1/2,1/2)
points to a dominant intra-plane ferromagnetic diagonal inter-dimer coupling. 
Finally, Hund's rule
exchange and the staggered rotation of the Ir-O octahedra
are known to induce anisotropic pseudo-dipolar and
Dzyaloshinskii-Moriya interactions~\cite{Jackeli2009}. We therefore consider the
Hamiltonian
\begin{align}
 H=&J\sum_{<i,j>,l}\Big[\cos(2\theta)\mathbf{S}_{li}\cdot\mathbf{S
}_{lj} +2\sin^{2}(\theta)S^{z}_{li}S^{z}_{lj}+ \nonumber \\
 &-\epsilon_{i}\epsilon_{l}\sin(2\theta)(\mathbf{S}_{li}
\times\mathbf{S}_{lj})\cdot\hat{e}_{z}\Big]+J_c\sum_{i}\mathbf{S}_{1i}\cdot\mathbf{S}_{2i}\nonumber \\
%&++J_{2c}\sum_{<i,j>}
%\mathbf { S }_{1i}\cdot\mathbf{S}_{2j}+ \nonumber \\
&+J_2\sum_{\ll i,j\gg,l}\mathbf{S}_{li}\cdot\mathbf{S}_{lj}
+J_3\sum_{ \lll i,j\ggg,l}\mathbf{S } _ { li } \cdot\mathbf{S}_{lj},
\label{Hextended}
\end{align}
where, in agreement with \srsl, a third neighbor in-plane coupling has been included\cite{Kim2012c}. The naming
convention for the exchange constants is indicated in Fig.\ \ref{fig1}(a).
In principle, due to the staggered rotation of IrO$_6$ octahedra, all bonds
connecting opposite sublattices have anisotropic exchange contributions, but one
can gauge away some of them\cite{Aharony1992}, e.g. that on the interlayer coupling $J_c$, which we
have chosen to do.
% leading out-of-plane anisotropy. In
%the latter case only in-plane interactions connecting same sub-lattice are
%anisotropic. For simplicity we keep only the leading anisotropic term on $J$,
%neglecting it on $J_{2c}$. 
In single layer \srsl\ the angle $\theta$ can be
inferred directly from the canting of the in-plane ordered
moment~\cite{Boseggia2013}. In \srbl\ the moments order along the c-axis, and
$\theta$ is a just a measure of the relative strength of the anisotropic interactions.

\subsection{Bond-operator mean-field method}

This model has been treated using bond-operator mean-field theory introduced by Sachdev and Bhatt \cite{Sachdev1990}, which has proved to be an accurate method for many bilayer spin systems and was especially applied to spin ladder systems \cite{Gopalan1994,Sushkov1998,Bouillot2011}. In this formalism, first of all, a close-packed dimerization of the lattice is chosen. In the case of our bilayer system, we designate interlayer bonds on the sites $(1,i)-(2,i)$, where the first coordinate refers to the layer and the second one to the position inside the layer. The four states $\ket{\uparrow\downarrow}$, $\ket{\downarrow\uparrow}$, $\ket{\uparrow\uparrow}$ and $\ket{\downarrow\downarrow}$ on each rung $i$ in the Hilbert space can be combined to form one singlet state $\ket{s_i}$ and three triplet states $\ket{t^{\alpha}_i}$, $\alpha{=}x,y,z$. To do so, singlet and triplet creation operators that create the states out of the vacuum $\ket{0}$ are introduced
\begin{align}
 \ket{s_i}\equiv& s^{\dag}_i\ket{0}=\frac{1}{\sqrt{2}}(\ket{\uparrow\downarrow}-\ket{\downarrow\uparrow}),
\nonumber\\
        \ket{t^{x}_i}\equiv& t^{x,\dag}_i\ket{0}=-\frac{1}{\sqrt{2}}(\ket{\uparrow\uparrow}-\ket{\downarrow\downarrow}),
\nonumber\\
        \ket{t^{y}_i}\equiv& t^{y,\dag}_i\ket{0}=\frac{\rm{i}}{\sqrt{2}}(\ket{\uparrow\uparrow}+\ket{\downarrow\downarrow}),
\nonumber\\
        \ket{t^{z}_i}\equiv& t^{z,\dag}_i\ket{0}=\frac{1}{\sqrt{2}}(\ket{\uparrow\downarrow}+\ket{\downarrow\uparrow}).
\end{align}
On each rung, the operators must fulfill a local hard-core constraint $s^{\dag}_is_i+\sum_{\alpha}t^{\alpha,\dag}_it^{\alpha}_i=1,~\alpha=x,y,z$. The action of the spin operators on the singlet and triplet states is then
equivalent to
\begin{align}
 \mathbf{S}_{1,i}^{\alpha}&=\frac{1}{2}\left(s^{\dag}_it^{\alpha}_i+t^{\alpha,\dag}_is^{\phantom{\dag}}_i-{\rm i}
\sum_{\beta,\gamma}\epsilon_{ \alpha\beta\gamma}t^{\beta,\dag}_it^{\gamma}_i\right), \nonumber \\
 \mathbf{S}_{2,i}^{\alpha}&=\frac{1}{2}\left(-s^{\dag}_it^{\alpha}_i-t^{\alpha,\dag}_is^{\phantom{\dag}}_i-{\rm i}
\sum_{\beta,\gamma}\epsilon_{\alpha\beta\gamma}t^{\beta,\dag}_it^{\gamma}_i\right).
\end{align}

\subsubsection{Dispersion}

To implement the mean-field approximation, one has to make assumptions about the ground state. Experimentally, it is known that the system has easy $c$-axis collinear AF magnetic order. We therefore describe the ground state by a condensation of singlet and triplet $t^{z}$ operators, $\ket{GS}=\prod_{i}\tilde{s}^{\dag}_{i},\,\braket{\tilde{s}}\neq0$, where a new operator basis is defined through rotation with a rotation angle $\chi$ adjusted to eliminate linear terms in the bosonic Hamiltonian, $4\left(J-J_2-J_3\right)\cos(2\chi)=J_c$:
\begin{align}
  \tilde{s}^{\dag}_{i}=&\cos(\chi)s^{\dag}_{i}-\epsilon_{i}\sin(\chi)t^{z,\dag}_{i},~\tilde{t}^{x,\dag}_{i}=t^{x,\dag}_{i}, \nonumber \\
  \tilde{t}^{y,\dag}_{i}=&t^{y,\dag}_{i},~\tilde{t}^{z,\dag}_{i}=\epsilon_{i}\sin(\chi)s^{\dag}_{i}+\cos(\chi)t^{z,\dag}_{i}.
\end{align}
Here, $\epsilon_i=e^{i\mathbf{Q}\cdot\mathbf{R}_i}$ with $\mathbf{q}=(\pi,\pi)$. Inserting this representation into the Hamiltonian of Eq.(1), some simplifications are possible: three-triplet terms have no contribution since they change sign under reflection along a plane perpendicular to the $c$-axis and passing through the centre of the rungs. Therefore, they vanish when taking the expectation value \cite{Gopalan1994}. Additionally, quartic triplet terms can be neglected due to their marginal effect on the results. This corresponds to ignoring triplet-triplet interactions.

The resulting Hamiltonian can be easily diagonalized with the help of a Bogoliubov transformation. We obtain a longitudinal mode $\omega_{\mathbf{q},z}$ and a two-fold degenerate transverse mode
$\omega_{\mathbf{q},x}=\omega_{\mathbf{q},y}$
\begin{equation}
        \omega_{\mathbf{q},\alpha}=\sqrt{A_{\mathbf{q},\alpha}^{2}-|B_{\mathbf{q},\alpha}|^{2}}, \label{BOMFdisp}
\end{equation}
with $\alpha=x,y,z$ and
\begin{widetext}
\begin{align}
  A_{\mathbf{q},z}=&4J\left[\sin^{2}(2\chi)\left(1-\frac{J_2}{J}-\frac{J_3}{J}\right)+\frac{J_c}{4J}\cos(2\chi)\right] +\frac{J}{2}\left[\cos^{2}(2\chi)\gamma_{\mathbf{q}}+\frac{J_2}{J}\delta_{\mathbf{q}}+\frac{J_3}{J}\varphi_{\mathbf{q}}\right], \nonumber\\
 B_{\mathbf{q},z}=&\frac{J}{2}\left[\cos^{2}(2\chi)\gamma_{\mathbf{q}}+\frac{J_2}{J}\delta_{\mathbf{q}}+\frac{J_3}{J}\varphi_{\mathbf{q}}\right], \nonumber \\
 A_{\mathbf{q},\tau}=&2J\left[\frac{J_c}{2J}\cos^{2}(\chi)+\sin^{2}(2\chi)\left(1-\frac{J_2}{J}-\frac{J_3}{J}\right)\right]+\frac{J}{2}\left[\cos(2\theta)\cos(2\chi)\right]\gamma_{\mathbf{q}}+\frac{J_2}{2}\delta_{\mathbf{q}}+\frac{J_3}{2}\varphi_{\mathbf{q}}, \nonumber\\
 B_{\mathbf{q},\tau}=&\frac{J}{2}\left[\cos(2\theta)-\rm{i}\sin(2\theta)\sin(2\chi)\right]\gamma_{\mathbf{q}}+\frac{J_2}{2}\cos(2\chi)\delta_{\mathbf{q}}+\frac{J_3}{2}\cos(2\chi)\varphi_{\mathbf{q}},
\label{eqABw}
\end{align}
\end{widetext}
where $\delta_{\mathbf{q}}=2(\cos(q_{x}+q_{y})+\cos(q_{x}-q_{y}))$, 
$\gamma_{\mathbf{q}}=2(\cos q_{x}+\cos q_{y})$,
$\varphi_{\mathbf{q}}=2(\cos 2q_{x}+\cos 2q_{y})$, and $\tau=x,y$.

As long as $\chi>0$, the gap of the transverse mode at $\mathbf{q}=(\pi,\pi)$ is given by
\begin{equation}
 \Delta E_S=\sqrt{J J_c}\sqrt{\frac{4J_2+4J_3-4J-J_c}{J_2+J_3-J}}\sin(2\theta).
 \label{bandgap}
\end{equation}

\begin{figure}
   \includegraphics[width=0.75\columnwidth]{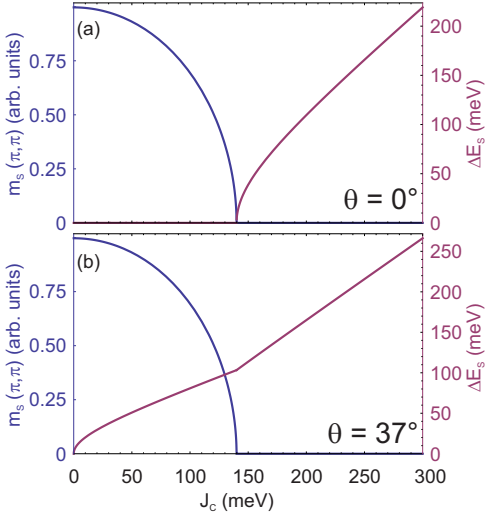}
   \caption{Staggered magnetization $m_S$ and magnetic gap $\Delta_S$ as a function of the interlayer coupling $J_c$ for (a) the isotropic case ($\theta=0^{\circ}$) and (b) the anisotropic case ($\theta=37^{\circ}$). The other coupling constants are equal to $J=26$~meV, $J_2=-15$~meV and $J_3=6$~meV. In both cases, the rotation angle $\chi$ vanishes for $J_c\geq140$~meV, bringing the system into the disordered phase.}\label{Jc-Bandgap}

\end{figure}

In the absence of anisotropy ($\theta=0$), the model has two phases: i) An ordered phase with a finite staggered moment as long as $\chi>0$. There is a Goldstone mode at ($\pi,\pi$), and this phase is thus gapless (the gap of Eq.(\ref{bandgap}) vanishes for $\theta=0$); ii) A gapped, disordered phase with no staggered magnetization when $\chi=0$. The gap closes at the transition. This is illustrated in Fig.~\ref{Jc-Bandgap}(a). In the presence of anisotropy ($\theta\neq0$), as in the bilayer iridate system, there is still a phase transition at which the staggered magnetization disappears, but the transverse mode acquires a gap in the ordered phase, as emphasized in Eq.(\ref{bandgap}). Accordingly, in the disordered phase, the gap does not vanish at the transition. This is illustrated in Fig.~\ref{Jc-Bandgap}(b). In view of its properties, a reduced but still significant staggered magnetization and a large gap, we think that the compound \srbl\ lies in the intermediate range, on the left of the transition.

\subsubsection{Intensity}
The dynamical scattering function $I^{\beta}_{q_{c}}(\mathbf{q},\omega)$, $\beta=x,y,z$ is proportional to the spin-spin correlation function $S^{\beta\beta}_{q_{c}}(\mathbf{q},\omega)$ given by
\begin{align}
S^{\beta\beta}_{q_{c}}(\mathbf{q},\omega)=\frac{2\pi}{L}\sum_{\lambda}|\bra{\lambda}S^{\beta}_{q_{c}}(\mathbf{q})\ket{0}|^{2}
\delta(\omega+\omega_{0}-\omega_{\lambda}) \label{intens}
\end{align}
at zero temperature with the symmetric ($q_{c}=0$) and antisymmetric ($q_{c}=\pi$) rung operators defined in real space as $S^{\beta}_{j,q_{c}}:=S^{\beta}_{1,j}\pm S^{\beta}_{2,j}$. The excited states $\ket{\lambda}$ are the lowest excited states with only one triplet excitation and an energy $\omega_{\lambda}$. The energy of the ground state $\ket{0}$ is $\omega_{0}$. Here, $\mathbf{q}$ is a two-dimensional vector describing the in-plane momentum transfer and $q_{c}$ is the phase of the out-of-plane momentum of the excitations. Note that the relation between $q_c$ and the out-of-plane momentum $L$ measured in experiment is given by $q_c =2\pi Ld/c (\text{mod.} 2\pi)$, where $c=20.8$~$\AA$ denotes the lattice parameter perpendicular to the bilayer, and $d=5.1$~$\AA$ is the intralayer distance.

Evaluating Eq.(\ref{intens}) using the bond-operator method, we obtain expressions for the symmetric and antisymmetric part of the transverse ($I^T_{q_c}$) and longitudinal ($I^L_{q_c}$) intensities. Since the transverse dispersion branch is twofold degenerate, its intensity is the sum of the intensities of the two degenerate modes $I^T_{q_c}=I^x_{q_c}+I^y_{q_c}$. This
leads to the integrated intensities
\begin{align}
&I^{T}_{0}(\mathbf{q})\propto2\sin^{2}(\chi)\frac{A_{-\mathbf{q}+\pi,\tau}+\text{Re}(B_{-\mathbf{q}+\pi,\tau})}{\omega_{
-\mathbf{ q}+\pi,\tau}}, \nonumber \\
&I^{T}_{\pi}(\mathbf{q})\propto2\cos^{2}(\chi)\frac{A_{\mathbf{q},\tau}-\text{Re}(B_{\mathbf{q},\tau})}{\omega_{\mathbf
{q},\tau}}, \nonumber \\ &I^{L}_{0}(\mathbf{q})=0, \nonumber \\
&I^{L}_{\pi}(\mathbf{q})\propto\cos^{2}(2\chi)\frac{A_{\mathbf{q},z}-\text{Re}(B_{\mathbf{q},z})}{\omega_{
\mathbf{q},z}}.
\end{align}
Remarkably, the longitudinal intensity has no symmetric part. 

For a three-dimensional stacking of bilayers, the dynamical scattering function is a linear combination of the symmetric and antisymmetric ones, with coefficients which depend on the out-of-plane  momentum according to:
\begin{align}
I&^{T/L}(\mathbf{q},q_{c},\omega) \nonumber \\
 &\propto
\cos(\frac{q_{c}}{2})^{2}I^{T/L}_{0}(\mathbf{q},\omega)+\sin(\frac{q_{c}}{2})^{2}I^{T/L}_{\pi}(\mathbf{q},\omega).
\label{Ints}
\end{align}

\subsubsection{Staggered magnetization}

Finally, the antiferromagnetic structure of \srbl\ can be theoretically reproduced by calculating the staggered magnetization $m_S(\mathbf{q})$ with $\mathbf{q}=(\pi,\pi)$. 
Since bond-operator theory breaks rotational symmetry in the antiferromagnetically ordered phase, the staggered magnetization can be obtained via the ground state expectation value at zero temperature of the local spin operator projected along the $c$-axis
\begin{equation}
 m_S(\mathbf{q})=g\mu_B\left|\bra{0}S^z_l(\mathbf{q})\ket{0}\right|
\label{MQ}
\end{equation}
where $l=1$ or $2$ refers to the layer, $\mu_B$ is the Bohr magneton and $g=2$ the electron spin g-factor. Expressing Eq.(\ref{MQ}) via bond-operator method, we obtain
\begin{equation}
 m_S(\mathbf{q})=\frac{g\mu_B}{2}\left|1-\alpha\right|\left|\sin(2\chi)\right|
 \label{moment}
\end{equation}
where $\alpha=\frac{1}{\pi^2}\int^\pi_0 d\text{k}_x\int^\pi_0 d\text{k}_y |v^z_\mathbf{k}|^2$ 
keeps track of the reduction of the staggered magnetization due to quantum fluctuations. In this expression, $v^z_\mathbf{k}$ is one of the coefficients of the Bogoliubov transformation $u^z_\mathbf{q}$ and $v^z_\mathbf{q}$, which can be expressed in terms 
of the dispersion given in Eq.(\ref{BOMFdisp}) as
\begin{equation}
 u^z_\mathbf{q}=\sqrt{\frac{A_{\mathbf{q},z}+\omega_{\mathbf{q},z}}{2\omega_{\mathbf{q},z}}},~~v^z_\mathbf{q}=\frac{B_{\mathbf{q},z}}{\sqrt{2\omega_{\mathbf{q},z}(A_{\mathbf{q},z}+\omega_{\mathbf{q},z})}}.
\end{equation}

\subsection{Discussion}

The results of the bond-operator mean-field theory applied to a bilayer with 
significant interlayer coupling and some anisotropy can be summarized as follows:
i) there is a main band of transverse excitations whose dispersion reflects to a
large extent the Fourier transform of the inter-dimer coupling; ii) the
anisotropic couplings induce a longitudinal excitation with a well-defined
dispersion whose energy is comparable to that of the main excitation band (by
contrast in the SU(2)-symmetric case longitudinal excitations only consist of a
continuum of two-magnon excitations); iii) the intensity of the transverse
excitation is larger than that of the longitudinal one, and it peaks at
(1/2,1/2), with a ratio to the smallest intensity typically in the range 1.5-3.

Remarkably, all of the characteristics of our model are qualitatively
consistent with the experimental results. Features B and D are ascribed to
transverse and longitudinal excitations, respectively. Feature C is the lower
boundary of the two-magnon continuum, which will be dominated by the dispersion
of B shifted up in energy by the 85~meV gap, as indicated by the dashed lines in
Figs.\ \ref{fig2}(a) and \ref{fig3}(c). In other words, regardless of the
details of the model, the assumption that the system can be described as coupled
dimers with some anisotropy leads to predictions that are supported by our new
RIXS data. To go beyond this qualitative observation, and in the absence of
strong constraints  provided by e.g. {\it ab initio} calculations, we optimized
the parameter set that agrees best with the dispersion of both modes and their 
intensity: $J=26$~meV, $J_2=-15$~meV, $J_3=6$~meV, $J_c=90$~meV, and
$\theta=37^{\circ}$. The dispersion curves calculated for this parameter set are
plotted in Fig.\ \ref{fig2}(a) and Fig.\ \ref{fig3}(c).

The only aspect of the dispersion that is not accurately reproduced by our theory
is the fact that the longitudinal mode seems to lie below the transverse one at ($\pi,0$). We note however that, according to a very recent improvement of the bond-order mean-field theory in the context of a $1/d$ expansion\cite{Joshi_I_2015,Joshi_II_2015}, the main effect of quantum fluctuations for a simple bilayer model in $d=2$ is to modify the spectrum of the longitudinal mode except at ($0,0$), while leaving the transverse mode unaffected. So this discrepancy is likely an artefact of the bond-order mean-field theory. 

The exchange pathways included in our Hamiltonian match those that would be
obtained by projecting a Hubbard model to fourth order~\cite{DallaPiazza2012}, except for the cyclic 4-spin terms that we have omitted for simplicity. In cuprates the 4-spin terms result in a zone boundary dispersion corresponding to an effective ferromagnetic $J_2$~\cite{Guarise2010}; a similar effect may explain the ferromagnetic next-nearest intra-layer coupling $J_2$ reported in \srsl\ \cite{Kim2012c} and also found here in Sr$_3$Ir$_2$O$_7$. The strong inter-layer coupling $J_c=90$~meV is qualitatively consistent with the very large bilayer splitting in the band structure measured by ARPES \cite{Moreschini2014}. Concerning the ratio of inter- to intra-layer coupling $J_c/J=3.5$, we note that in the SU(2) symmetric case, linear bond-operator theory overestimates the quantum critical ratio $J_c/J=4$ ~\cite{Sommer2001} whereas numerical methods place it at 2.51~\cite{Sandvik1994}. It is therefore plausible that treating our Hamiltonian to higher order would decrease the extracted ratio correspondingly by increasing $J$.

With this set of parameters, adjusted to fit the dispersion only, the
intensities of the transverse and longitudinal modes (see Fig.\ \ref{fig4}(b))
are in reasonable agreement with the experimental results: the response peaks
around (1/2,1/2), and the transverse mode is more intense across the whole
Brillouin zone than the longitudinal one. However, our theory appears to
overestimate the intensity of the transverse mode relative to the longitudinal
one. 

\begin{figure}
\centering
  \vspace{2ex}
  \includegraphics[width=0.75\columnwidth]{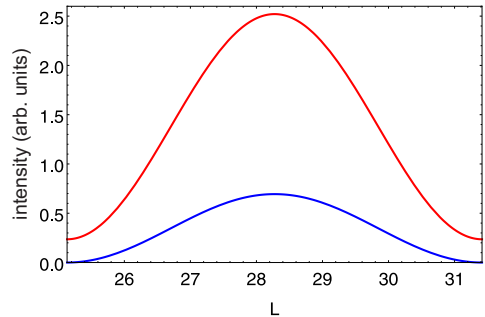}
 \caption{Dependence of the longitudinal (blue) and transverse (red)
intensities on the out-of-plane momentum $L$ at the reciprocal space point
$(\pi,\pi)$.}
\label{Qc-plot-Pi-Pi}
\end{figure}

The $L$ dependence of the RIXS spectrum has also been calculated (see
Fig.\ \ref{Qc-plot-Pi-Pi}) and is qualitatively in agreement with the data shown in Fig.\ \ref{fig2}(b). Indeed, the intensities vary periodically with $q_c$, reaching their maximum at $q_c=\pi$ ($L=28.3$) (Fig.\ref{Qc-plot-Pi-Pi}). While $I^T$ is finite for all momenta, $I^L$ vanishes when $q_c$ is a multiple of $2\pi$. In particular, at $L=25$, $q_c/2\pi=6.1$, i.e. $q_c$ is nearly a multiple of $2\pi$, and the longitudinal intensity nearly vanishes, in agreement with the theoretical prediction. We believe that this explains why the D feature has not been detected in the previous experiment.

Finally, inserting the values for the coupling constants and the rotation angle
$\chi=\frac{1}{2}\arccos\left(\frac{J_c}{4(J-J_2-J_3)}\right)\approx25^{\circ}$ into 
Eq.(\ref{moment}), we get a staggered magnetization of
\begin{equation}
 \frac{g}{2}\left|\sin(2\chi)\right|\left|1-\alpha\right|\mu_B=0.76\,\mu_B.
\end{equation}
In this formalism, the AF order is a consequence of the non-zero value of the
rotation angle $\chi$, which, in the groundstate, mixes singlets and triplets on a dimer. The correction due to quantum fluctuations is very small
($\alpha\approx0.006$) because the spectrum has a large gap. As compared to the experimental value of about $0.52\pm0.08\mu_B$ \cite{Dhital2012}, the
theoretical value is somewhat larger. It would be easy to get a smaller moment
by choosing  a smaller rotation angle $\chi$ at the expense of the quality of
the fit of the dispersion. However, we have not attempted to do it since the 
bond-operator mean-field theory should not be considered as quantitatively
accurate. 

\section{Conclusion}

Putting our results in perspective, we first recall that bilayer systems with
SU(2)-symmetry have a quantum critical point (QCP) as a function of the
interlayer coupling between an antiferromagnetic phase and a gapped phase (Fig.\ \ref{fig1}(d)). When anisotropy in spin space becomes important, as in
systems with strong spin-orbit coupling such as \srbl, a new paradigm arises
where the quantum critical point is replaced by a simple transition between a gapped antiferromagnet and a gapped quantum dimer system, as sketched in Fig.\ \ref{fig1}(e). Considering the failure of linear spin-wave theory to explain the new mode reported in this paper together with the presence of a significant staggered magnetization, we are led to the conclusion that the system lies in the intermediate regime, on the left of the point where the antiferromagnetic order disappears, but with excitations of dominantly quantum dimer character. It will be rewarding to test this conclusion with more sophisticated theoretical approaches that could allow one to reach a fully quantitative agreement with experiments.

\begin{acknowledgments}
We gratefully thank C. Henriquet and R. Verbeni for technical assistance during the experiements and B. Normand, O. Syljuaasen and B. Dalla Piazza for
insightful discussions. The work in Lausanne was supported by the Swiss National Science Foundation and its Sinergia network Mott Physics Beyond the Heisenberg Model, in London by the EPSRC and in W\"{u}rzburg by the ERC starters grant TOPOLECTRICS under ERC-StG-Thomale-336012.
\end{acknowledgments}

%Bibliography
\bibliographystyle{apsrev4-1}
%\bibliography{Sr327_final}

\end{document}